\documentclass[prb,twocolumn,showpacs,preprintnumbers,amsmath,amssymb]{revtex4-1}

\usepackage{graphicx}% Include figure files
\usepackage{bm}% bold math
\usepackage{graphicx}
\usepackage{epsfig}
\include{graphic}

\begin{document}
\preprint{APS/123-QED}

\title{Rapid microwave phase detection based on a solid state spiontronic device}
\author{B. M. Yao$^{1,2}$, L. Fu$^{1,2}$, X. S. Chen$^2$, W. Lu$^2$, L. H. Bai$^1$, Y. S. Gui$^1$ and C.-M. Hu$^1$}

\affiliation{$^{1}$Department of Physics and Astronomy, University
of Manitoba, Winnipeg, Canada R3T 2N2}

\affiliation{$^{2}$National Laboratory for Infrared Physics,
Chinese Academy of Science, Shanghai 200083, People's Republic of
China}

\date{\today}

\begin{abstract}
A technique for rapidly detecting microwave phase has been developed which uses a spintronic device that can directly rectify microwave fields into a dc voltage signal. Use of a voltage-controlled phase shifter enables the development of a spintronic device that can simultaneously "read" the magnitude and phase of incident continuous-wave (CW) microwaves when combined with a lock-in amplifier. As an example of many possible practical applications of this device, the resonance phase in a complementary electric inductive-capacitive (CELC) resonator has been characterized using a spintronic sensor based on a magnetic tunnel junction (MTJ). This sensor device is not limited for use only with spintronic devices such as MTJs, but can also be used with semiconductor devices such as microwave detectors, and hence offers a useful alternative to existing microwave imaging and characterization technologies.
\end{abstract}



\maketitle

%\section{INTRODUCTION}
Converting dc currents into microwaves (and visa versa) within ferromagnetic materials and structures through spin dynamics has become an extremely interesting topic due to the fundamental physics involved as well as many possible technological applications\cite{Kiselev2003, Houssameddine2008, Houssameddine2009, Tulapurkar_Spindiode, SRE, Wang2009, Ishibashi2010, Phase-Andre, Phase-Fan, Zhang2012, Cao2012, Fu2012}. The magnetic tunnel junction (MTJ) is of particular interest among ferromagnetic structures due to the fact that it is able to convert dc currents into microwaves, acting as a microwave source\cite{Kiselev2003, Houssameddine2008, Houssameddine2009}, in addition to being able to convert microwave energy into a dc current, thus acting as a microwave sensor\cite{Tulapurkar_Spindiode, Wang2009, Ishibashi2010, Zhang2012}. Recently, the practical application of an MTJ based microwave sensor has been reported for microwave imaging based on the dielectric contrast of materials\cite{Fu2012}. Compared to traditional antennas used in microwave imaging, spintronic sensors have the advantages of broad bandwidth, small size, an easily detected dc output, and the capability to detect both the electric and magnetic fields of a microwave signal, making them a very promising alternative to existing microwave imaging technology.

Hindering the development of spintronic sensors (as well as any other kind of microwave sensor) is the phase measurement of continuous wave (CW) microwaves, as the microwave phase contains information crucial for a proper estimation of electric permittivity and conductivity profiles. In traditional time-domain microwave sensing systems the microwave phase is measured using expensive and complicated equipment such as vector network analysers (VNAs), which permit both the in-phase and quadrature components of a microwave to be detected by an antenna. Spintronic devices, despite having the ability to detect microwave phase through the interference principal\cite{Phase-Andre, Phase-Fan, Cao2012}, take far longer to complete a phase measurement when compared to a VNA system due to the fact that one has to turn the phase shifter and perform a computer fitting of the measured alternating interference fringes of rectified voltage in order to determine microwave phase\cite{Phase-Andre, Phase-Fan, Cao2012}.

In this letter we demonstrate an advancement in using spintronic devices for microwave measurement, permitting the magnitude and phase of a CW microwave to be simultaneously measured using an MTJ microwave sensor. This advancement allows a lock-in amplifier to perform real-time microwave measurement through spintronic devices, analogous to measurements taken by VNAs through antennas.

\begin{figure} [t]
\begin{center}
\epsfig{file=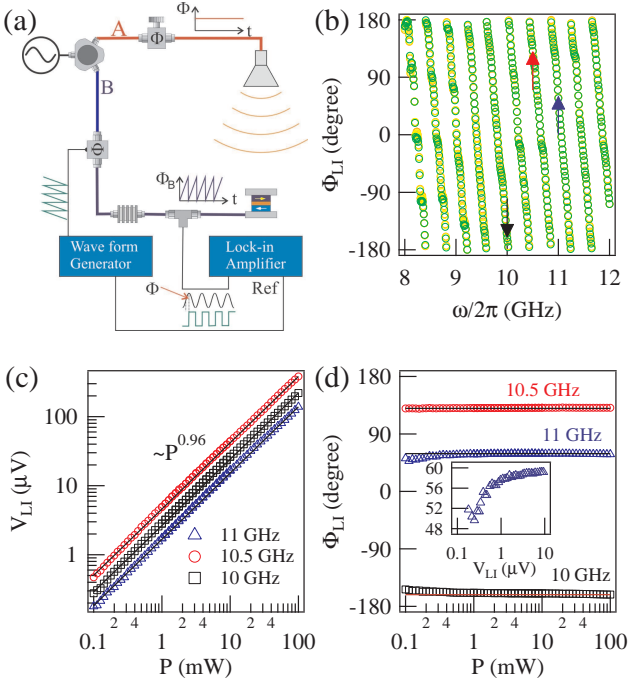,width=8.5 cm} \caption{(color online).  (a) Interferometry experimental set-up, through which the microwave signal is coherently split into two beams and finally coupled at the sensor (a magnetic tunnel junction). The usage of a voltage controlled phase shifter inserted into path B enables the detection of microwave magnitude and phase using an MTJ microwave sensor. (b) Fixing the mechanical phase shifter to be zero, the measured lock-in phase $\Phi_{LI}$ plotted as a function of microwave frequency for several microwave powers $P$=10 (orange circles), 15 (yellow circles), and 20 (green circles) dBm. Black, red and blue arrows indicate the lock-in phase $\Phi_{LI}$ detected at $\omega/2\pi=$ 10, 10.5 and 11 GHz, respectively. (c) Lock-in magnitude $V_{LI}$ (symbols) follows a $\propto{P^{0.96}}$ relation (solid lines), where $P$ is microwave power.  (d) Lock-in phase $\Phi_{LI}$ is not sensitive to microwave power. Deviation of $\Phi_{LI}$ appears when $V_{LI}<1$ $\mu$V.
}\label{fig1}
\end{center}
\end{figure}

To rapidly detect microwave phase through the use of spintronic devices, the measurement system detailed in Fig. \ref{fig1} (a) has been developed. This system uses an MTJ microwave sensor to rectify microwave fields in the absence of an external magnetic field using the Seebeck rectification effect\cite{Zhang2012}. Under microwave radiation, the inner components of the MTJ are subjected to Joule heating due to the microwave current generated within the structure, raising its internal temperature. As the internal structure of the MTJ is asymmetric, this temperature increase produces a temperature gradient across the MgO barrier layer; this gradient produces Seebeck rectification within the MTJ\cite{Zhang2012}.

The measurement system includes a broadband microwave generator (Agilent E8257D) from which the output electromagnetic field is coherently split into two parts by a microwave power divider, each part travels along a different path before both paths finally couple at the MTJ microwave sensor. Path A includes an SMA to waveguide adapter, a horn antenna,  and several coaxial cables. To verify the accuracy of the phase shift measured by this approach, a broadband mechanical adjustable phase shifter is inserted into path A, which can continuously adjust microwave phase, $\Phi$, during experiments to within 2$^\circ$ of accuracy. Path B includes a voltage-controlled phase shifter, an adjustable attenuator for balancing the strength of the two beams, a bias tee for separating rf and low frequency signals, an MTJ microwave sensor, and several coaxial cables.

What makes this experimental set-up such a significant technical advancement is its use of a voltage-controlled phase shifter, which has the phase delay depend linearly on the voltage bias. This voltage-controlled phase shifter is connected to a function generator which produces a sawtooth wave at a frequency of $\omega_V$ ($\omega_V{\ll}\omega$), where $\omega$ is the microwave frequency. The range of the voltage bias is set so that the phase delay can vary from 0 to 360$^\circ$; i.e. the microwave phase for path B is $\Phi_B(t)=\omega_Vt+\Phi_0$ where $\Phi_0$ is the initial microwave phase of path B. Thus a voltage generated across the MTJ (after eliminating terms which are phase insensitive after time averaging) is given by

\begin{eqnarray}
V\sim{}e_T\cos(\omega{t}+\Phi)\cdot{}e_R\cos(\omega{t}+\omega_Vt+\Phi_0) \nonumber \\
=[e_Te_R\cos(2\omega{}t+\omega_Vt+\Phi+\Phi_0) \nonumber \\
+e_Te_R\cos(\omega_Vt+\Phi_0-\Phi)]/2,
\label{Eq:phasemath}
\end{eqnarray}

\noindent Where $e_T$ and $e_R$ are the magnitudes of the microwave electric fields for beams A and B, respectively. Equation (\ref{Eq:phasemath}) produces a second harmonic microwave signal with a frequency of about $2\omega$ as well as a low frequency signal of the form $\cos(\omega_Vt+\Phi_0-\Phi)$. This latter term carries the phase information of the microwave, $\Phi$, which is what we want to measure. Triggered by a square wave synchronized to the function generator output, a SR830 DSP (digital signal processing) lock-in amplifier is able to detect both the in-phase and quadrature components of the voltage across the MTJ varying with a frequency of $\omega_V$, allowing the microwave phase shift, $\Phi-\Phi_0$, to be detected.

To verify that the magnitude, $V_{LI}$, and phase, $\Phi_{LI}$, detected by the MTJ sensor respectively correlate to microwave power, $P$, and phase, $\Phi$, we perform an experiment where the mechanical adjustable phase shifter is set to zero ($\Phi=0$) and the lock-in phase is measured as a function of microwave frequency. The results are plotted in Fig. \ref{fig1}(b); $\Phi_{LI}$ shows a periodic linear dependence with frequency, which is expected since given a travel distance difference $d$ between the two beams of the system, the phase difference between them is $\Delta{\Phi}=(d{}f/\nu)\times360^\circ$, where $\nu$ and $f$ are the speed and frequency of the microwave. Note that the lock-in amplifier measures $\Delta\Phi$ with a modulus of $360^\circ$ over a range from -180$^\circ$ to 180$^\circ$, this results in the zigzag behaviour of $\Phi_{LI}$. As shown in Fig. \ref{fig1}(b), the lock-in phase appears insensitive to microwave power; the orange, yellow, and green circles in this figure represent $P$=10, 15, and 20 dBm, respectively.

In order to check the power dependence of measurements, microwave power was varied over three orders of magnitude, ranging from -10dBm (0.1mW) to 20dBm (100mW), for measurements at constant microwave frequencies; the power variance of $V_{LI}$ and $\Phi_{LI}$ for three frequencies, $\omega/2\pi=$ 10, 10.5 and 11 GHz [marked by the arrows in Fig. \ref{fig1}(b)], are shown in Fig. \ref{fig1}(c) and (d). The power dependence seen in Fig. \ref{fig1}(c) [$V_{LI}\propto{P^{0.96}}$] is in agreement with the expected relation of $V_{LI}\propto{e_Te_R}\propto{P}$. This agreement allows us to conclude that $V_{LI}$ indeed measures the power of the microwave. In contrast to the pronounced change seen in $V_{LI}$, the value of $\Phi_{LI}$ [symbols in Fig. \ref{fig1}(d)] is seen to be almost constant with respect to $P$, with a fluctuations of less than $\pm1^\circ$ when $V_{LI}>1$ $\mu$V; here the signal is at least 10 times greater than the background noise (which has an amplitude of about 100nV). The invariance of $\Phi_{LI}$ with respect to $P$ agrees with the expectations of Eq. (\ref{Eq:phasemath}). When $V_{LI}<1$ $\mu$V the experimental error in the value of $\Phi_{LI}$ becomes significant and can no longer be neglected; as seen in the inset of Fig. \ref{fig1}(d), the maximum deviation can be as high as 10$^\circ$ at $V_{LI}\sim200$nV, where the signal magnitude is near the background noise amplitude.

\begin{figure} [t]
\begin{center}
\epsfig{file=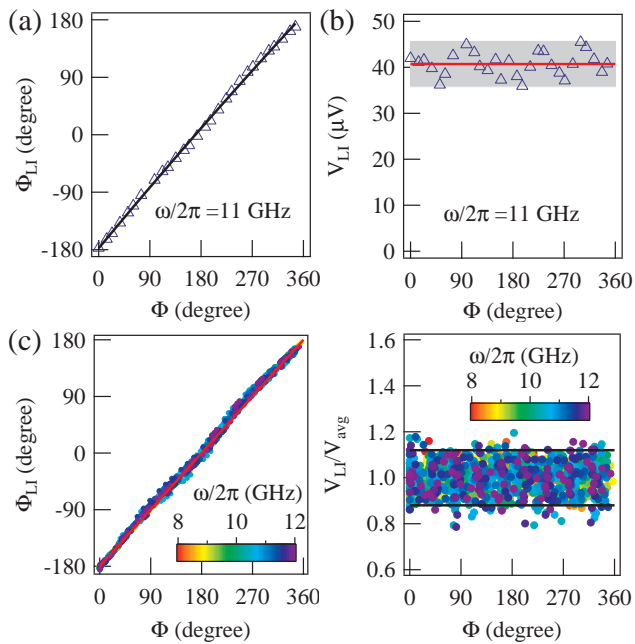,width=8.5 cm} \caption{(color online). The lock-in phase for each frequency is offset so that $\Phi_{LI}$ is $-180^\circ$ at $\Phi=0$ (the zero point of the mechanical phase shifter). (a) $\Phi_{LI}$ as a function of $\Phi$ at $\omega/2\pi$= 11 GHz. The value of $\Phi$ is adjusted by the mechanical phase shifter in path A as shown in Fig. \ref{fig1}(a). The solid line is the relation $\Phi_{LI}=\Phi-180^\circ$. (b) $V_{LI}$ as a function of $\Phi$ at $\omega/2\pi$= 11 GHz. The red line indicates  the average value of $V_{LI}$, $V_{avg}$, and the grey area indicates the area of average value with an errorbar of $\pm12\%$. (c) $\Phi_{LI}$ as a function of $\Phi$ for all frequencies ranging from 8 GHz to 12 GHz with a step size of 0.1 GHz. The solid line is the relation $\Phi_{LI}=\Phi-180^\circ$. (d) $V_{LI}/V_{avg}$ as a function of $\Phi$ for all frequencies. The two solid lines correspond $V_{LI}/V_{avg}$=0.88 and 1.12, respectively.
}\label{fig2}
\end{center}
\end{figure}

To further verify that by using a lock-in amplifier an MTJ sensor can be used to measure microwave phase, we perform a second experiment which involves adjusting the mechanical phase shifter in microwave path A and fixing the microwave power at 15dBm. This experiment should allow the dependence of $\Phi_{LI}$ on $\Phi$ to be determined. To facilitate a systematic comparison, the lock-in phase for each frequency is offset so that $\Phi_{LI}$ is $-180^\circ$ at $\Phi=0$ (the zero point of the mechanical phase shifter). After this offset, which is similar to the phase calibration process used for VNA measurements, the lock-in phases for all frequencies shown in Fig. \ref{fig1}(b) have baselines of $\Phi_{LI}=-180^\circ$.

Figure \ref{fig2}(a) illustrates the lock-in phases recorded (triangle symbols) while adjusting the mechanical phase shifter at $\omega/2\pi=11$GHz, which exactly follow the relation $\Phi=\Phi_{LI}+180^\circ$ (solid line). This result demonstrates that the lock-in amplifier can indeed directly measure microwave phase, confirming the predictions of Eq. (\ref{Eq:phasemath}). It should be noted that turning the phase, $\Phi$, introduces a maximum error of $\pm12\%$ into our measurements, as indicated by the grey area in Fig. \ref{fig2}(b).

To verify the stability of this measurement approach, we have measured about 30 individual $\Phi$ values at more than 40 microwave frequencies, and found $\Phi_{LI}$ to be very consistent for each frequency. As shown in  Fig. \ref{fig2}(c), $\Phi_{LI}$ (the colour of each symbol represents microwave frequency) neatly follows the relation $\Phi=\Phi_{LI}+180^\circ$ (solid line) for frequencies between 8GHz and 12GHz (the step size was 0.1GHz). We have also rescaled $V_{LI}$ for each of the various frequencies using their average values, $V_{avg}$, and plotted $V_{LI}/V_{avg}$ for each frequency in Fig. \ref{fig2}(d); nearly all data in this plot fell between $V_{LI}/V_{avg}=0.88$ and $V_{LI}/V_{avg}=1.12$. One source of the error in $V_{LI}$ comes from power loss within the mechanical phase shifter (approximate maximum of 1dBm); this only affects the microwave power in path A, so the maximum error in the value of $e_T$ is roughly $\pm6\%$ while the value of $e_R$ should not be affected. Therefore we can conclude that parasitic effects within our experimental set-up may introduce an error of about 6\%; this error may be caused by the impedance-mismatch of the system. Despite these errors, the scaling results shown in Fig. \ref{fig2}(c) and (d) confirm that the apparatus shown in Fig. \ref{fig1}(a) is able to fairly precisely detect both microwave magnitude and phase using an MTJ sensor.

\begin{figure} [t]
\begin{center}
\epsfig{file=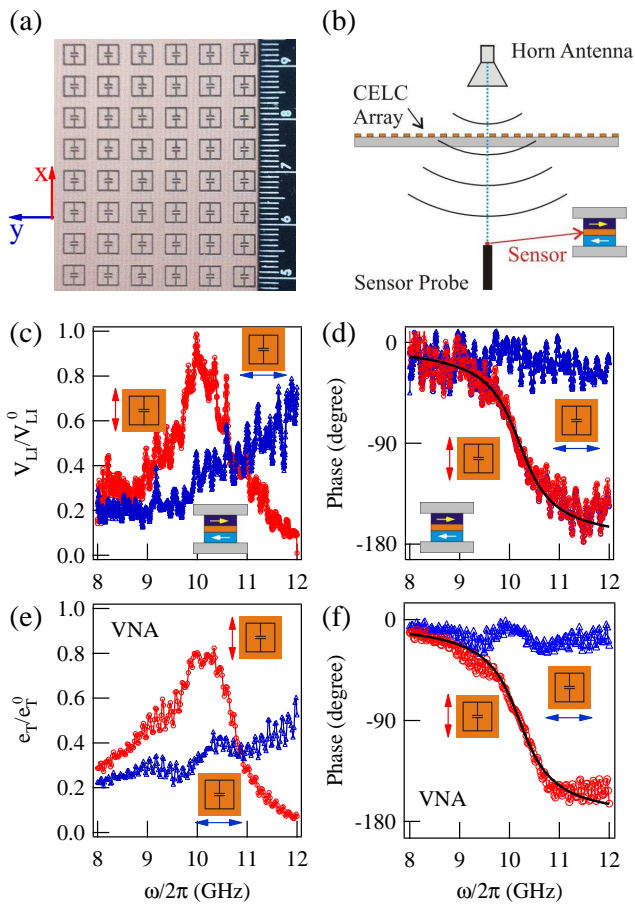,width=8.5 cm} \caption{(color online).  (a) The fabricated CELC array with a unit cell size of $6\times6$mm$^2$. (b) Diagram of the transmission measurement of the CELC array, where an X-band horn antenna acts as the transmitter and an MTJ acts as the receiver. (c) and (d) are, respectively, the magnitude and phase of the transmission coefficient measured by an MTJ using a lock-in amplifier. Red (blue) symbols correspond to the incident microwave magnetic field $\mathbf{h}=\hat{x}h$ ($\mathbf{h}=\hat{y}h$), which is perpendicular (parallel) to the gap of the CELC cell. The data is identical for measurements at different microwave powers(10, 15 and 20 dBm). The solid line in (d) is a calculation of the resonance phase using a resonance frequency of $\omega_0/2\pi=10.2$ GHz and a line width of $\Delta\omega/2\pi=0.5$ GHz. (e) and (f) are the same as (c) and (d), respectively, with measurements performed by an antenna connected to a VNA.
}\label{fig3}
\end{center}
\end{figure}

Next we apply our measurement system to characterizing complementary electric inductive-capacitive (CELC) resonators, which are of interest for both their metamaterial properties and their possible application in microwave computational imaging \cite{Hunt2013}. The complement to CELC resonators, electric field coupled inductive-capacitive (ELC) resonators, can be qualitatively described in terms of an equivalent circuit where a varying electric field drives the resonator at an eigenfrequency of $\omega_0=\sqrt{2/LC}$\cite{Schurig2006}. Although the equivalent circuit of a CELC resonator is difficult to accurately render, Babinet's Law states that CELC resonators should exhibit a purely magnetic response; this has been verified in Ref. \onlinecite{Hand2008}. The CELC resonator array used in our experiment was fabricated on a 1.5 mm thick FR4 board using a standard milling technique, and consists of a block of 30$\times$30 cells, each cell being $6\times6$ mm$^2$ in size. A diagram of the CELC resonator array is shown in Fig. \ref{fig3}(a) and the design of the CELC resonator cells follows the report\cite{Hand2008}. The transmission coefficient (namely its magnitude and phase) of the CELC resonator array was measured using the experimental set-up shown in Fig. \ref{fig3}(b), where an X-band horn antenna was used as a transmitter and an MTJ sensor acted as a receiver. The CELC resonator array was placed roughly halfway between the transmitter and receiver, with the MTJ sensor positioned to face the centre of the horn antenna.

Figure \ref{fig3}(c) shows the magnitude of the transmission coefficient, $V_{LI}/V_{LI}^0$, where $V_{LI}\propto{}e_Te_R$ is the lock-in magnitude measured with the CELC resonator array and $V_{LI}^0$ corresponds to a background measurement without the CELC resonator array. A resonance excitation (red symbols) appears when the microwave magnetic field $\mathbf{h}=\hat{x}h$ is along the x-direction [the coordinates are shown in Fig.\ref{fig3}(a)]. This resonance is driven by the microwave current in the y-direction induced by $\mathbf{h}$\cite{Hand2008}. Consistently, we do not observe resonance for $\mathbf{h}=\hat{y}h$ with the CELC array rotated 90$^\circ$, because in this case no microwave magnetic field is perpendicular to the gap complement of the CELC cells.

In order to have a deeper understanding of wave propagation through a CELC resonator array, the phase constant has also been measured. In agreement with the magnitude measurement, the phase constant shows an offset from 0 to 180$^\circ$ only for $\mathbf{h}=\hat{x}h$. This effect reflects the universal feature of resonators: the phase of the response always lags behind the phase of the driving field. Far from resonance ($\omega_0$), the response is in-phase with the driving field when the eigenfrequency of the resonator follows $\omega_0\gg\omega$, and the response is out-of-phase when $\omega_0\ll\omega$. From the measured resonance phase, $\Theta$, an eigenfrequency of $\omega_0/2\pi=10.2$ GHz and a line width of $\Delta\omega/2\pi=0.5$ GHz are calculated (solid line in Fig. \ref{fig3}(d)) from the relation $\tan(\Theta)=\Delta\omega/(\omega-\omega_0)$. For $\mathbf{h}=\hat{y}h$ there is no resonant excitation because there is no microwave magnetic field perpendicular to the CELC gap complement; therefore, a flat, constant phase curve which is almost in-phase with the driving field, is seen in Fig. \ref{fig3}(d).

By replacing the sensor probe with another X-band horn antenna and using an Agilent N5230C network analyzer to collect our data, we are able to compare the measurements performed by a lock-in amplifier with those from a VNA; Figure \ref{fig3}(e) and (f) show the results. The agreement between these approaches is remarkable. Additionally, the data acquisition time is only a few seconds for the measurement of the phase constant with the lock-in method, indicating real-time measurement of both microwave magnitude and phase using a spintronic sensor.

In summary, we report on a sensor approach for simultaneously measuring both the magnitude and phase of microwaves. This technique enables small integrated sensors to non-contactively and non-destructively characterize engineered structures such as metamaterials. This sensor based technique is broadband, does not employ complicated microwave electronics, and can easily be integrated, making it a promising alternative for use in various microwave applications. Without limiting its scope, this approach is applicable for use with any solid state sensor, including magnetic devices such as MTJs and semiconductor devices such as Schottky diodes and metal-insulator-metal diodes.

%\section{ACKNOWLEDGEMENTS}
We would like to thank Paul Hyde and Fuchun Xi for discussions.  This work has been funded by NSERC, CFI, CMC and URGP grants.  BMY was supported by the National Natural Science Foundation of China Grant No. 10990100 and No. 11128408.

%\clearpage

\end{document}